\documentclass[12pt,english,american]{article}
\usepackage{mathptmx}
\usepackage{helvet}
\usepackage{courier}
\usepackage[T1]{fontenc}
\usepackage[latin9]{inputenc}
\usepackage[a4paper]{geometry}
\usepackage{float}
\usepackage{graphicx}
\usepackage{setspace}
\makeatletter

\providecommand{\tabularnewline}{\\}

\newenvironment{lyxlist}[1]
{\begin{list}{}
{\settowidth{\labelwidth}{#1}
 \setlength{\leftmargin}{\labelwidth}
 \addtolength{\leftmargin}{\labelsep}
 }}
{\end{list}}

\makeatother

\usepackage{babel}

\begin{document}

\title{Quantum Secret Authentication Code}

\author{Tong-Xuan Wei, Tzonelih Hwang%
\thanks{corresponding author%
} ~and Chia-Wei Tsai }
\maketitle
\begin{abstract}
This study proposes a quantum secret authentication code for protecting
the integrity of secret quantum states. Since BB84\cite{key-1} was
first proposed, the eavesdropper detection strategy in almost all
quantum cryptographic protocols is based on the random sample discussion,
in which the probability of eavesdropper detection is depending on
the number of check qubits eavesdropped by the eavesdropper. Hence,
if the eavesdropper interferes only a few qubits of the transmitted
quantum sequence, then the detection probability will be very low.
 This study attempts to propose a quantum secret authentication code
to solve this problem. With the use of quantum secret authentication
code, not only is the probability of eavesdropper detection guaranteed
to be evenly distributed no matter how many qubits had been eavesdropped,
but also can the quantum transmission efficiency be highly enhanced.

\textbf{\emph{keywords: message authentication code, integrity, random
sampling}}
\end{abstract}

\section{Introduction}

In 1984, BB84, the first quantum cryptographic protocol was proposed.
For the next decade, the eavesdropper detection strategy, following
the idea of BB84, for almost all quantum cryptographic protocols is
based on the random sample discussion. In this eavesdropper detection
method, the check qubits are independent to the message qubits. By
scrambling the message qubits with the check qubits together, the
strategy assumes that an eavesdropper has no idea on discriminating
a check qubit from a message qubit. An attempt to eavesdrop the qubit
sequence by an outsider would only end up disturbing the check qubit
states. And, the only chance for an eavesdropper detection is when
the check qubits were altered {}``inadvertently'' by the eavesdropper.
Hence, the more check qubits altered by an eavesdropper, the higher
probability he/she is detected.

In order to increase the detection probability. The number of check
qubits should be large enough, because the eavesdropper detection
probability for interfering one check qubit is

\[
\frac{number\: of\: check\: qubits}{Tatal\: number\: of\: qubits}\times\frac{1}{4}\]
where the $\frac{number\: of\: check\: qubits}{Tatal\: number\: of\: qubits}$
represents the probability of the eavesdropper to choose a check qubit,
and $\frac{1}{4}$ is the probability that the measurement result
of that check qubit is different from what was expected due to the
interference of that eavesdropper. Most protocols suggest the number
of check qubits should be at least the same as that of the message
qubits transmitted \cite{key-4,key-5,key-6,key-7}.

However, there is a problem with the random sampling discussion strategy,
especially when an eavesdropper just attempts to eavesdrop only a
few qubits in the transmitted quantum sequence. In this situation,
some information might be revealed, and the integrity of the message
qubits might be jeopardized, but the eavesdropper detection probability
could still be very low. Most quantum cryptographic protocols ignore
this problem and conversely assume that the eavesdropper always attempts
to eavesdrop the entire sequence of qubits. Consequently, the conclusion,
the probability of eavesdropper detection can reach 1, is asserted.

This paper aims to design a quantum secret authentication code (QSAC)
to solve the above-mentioned problem. This QSAC guarantees the integrity
of the secret qubits transmitted from a sender to a receiver. Furthermore,
the intentionally designed avalanche effect guarantees the detection
of an eavesdropping of even a single qubit in the quantum sequence.

The rest of this paper is organized as follows. Section 2 presents
the proposed quantum secret authentication code and the security analysis.
Section 3 gives an application of the proposed QSAC. Finally, conclusions
are made in Section 4.

\section{Quantum secret authentication code}

In this section, a quantum secret authentication code (QSAC) is proposed
for a receiver to verify whether or not the received states are indeed
from the alleged sender and without being eavesdropped or modified.

\subsection{Notations}
\begin{lyxlist}{00.00.0000}
\item [{\emph{K}}] the pre-shared key between the sender and the receiver.
\item [{||}] concatenation.
\item [{$\left|M\right\rangle ^{i}$}] the \emph{i} th qubit in the quantum
sequence \emph{M}.
\item [{CNOT(\emph{a},\emph{b})}] Control-Not gate, where \emph{a} is the
control qubit and \emph{b} is the target qubit. If $a=b$, then performs
the identity operation \emph{I}.
\end{lyxlist}

\subsection{The avalanche effect}

A CNOT gate is applied here to create an avalanche effect in the QSAC.
Figure 1 demonstrates an avalanche effect of the QSAC. Assume that
there is a function \emph{F} with a sequence of quantum states as
the input and a sequence of entangled qubits as the output. The function
is composed of a sequence of \emph{CNOT} operations. For example,
let the input of the function be a three-qubit quantum state $\mbox{\ensuremath{\left|\psi\right\rangle }}_{123}$
and the output, $CNOT(q_{1},q_{2})$ $CNOT(q_{2},q_{3})CNOT(q_{3},q_{1})\left|\psi\right\rangle $.
Conversely, the inverse function $F^{-1}$ should be $CNOT$ $(q_{3},q_{1})CNOT(q_{2},q_{3})CNOT(q_{1},q_{2})\left|\psi\right\rangle $.
An avalanche effect is described under a communication model between
a sender and a receiver.  First, the sender inputs the state, $\left|000\right\rangle $
for example, to the function \emph{F }and subsequently, \emph{F} outputs
the state $\left|000\right\rangle $\emph{. }Then the sender sends
the qubits $\left|000\right\rangle $ to the receiver via a quantum
channel. During the time of transmission, if there is no eavesdropper,
then the same state $\left|000\right\rangle $ will be received and
recovered by the receiver, i.e., $\left|000\right\rangle \rightarrow F^{-1}\rightarrow\left|000\right\rangle $.
Conversely, assume the last qubit is altered by an outsider, i.e.,
the state received by the receiver becomes $\left|001\right\rangle $.
Consequently, the receiver inputs the received state $\left|001\right\rangle $
to $F^{-1}$, the inversion of the function \emph{F.} The state that
the receiver finally recovers is $\left|110\right\rangle $. Obviously,
one single qubit modification eventually causes more than one qubits
to change. %
\begin{figure}[H]
\centering{}\includegraphics[scale=0.3]{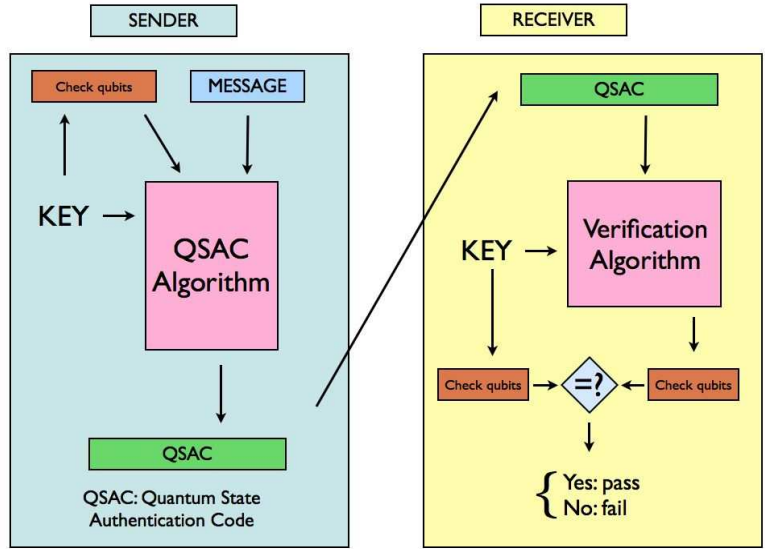}\caption{The avalanche
effect of the QSAC.}

\end{figure}

\subsection{The proposed QSAC}

\selectlanguage{english}%
\foreignlanguage{american}{Based on the technique described in Sec.
2.2, this section proposes a QSAC design that marries a pre-shared
key, \emph{K,} between a sender and a receiver to a message qubits,
$\left|M\right\rangle $ to be transmitted from the sender to the
receiver via a sequence of CNOT operations in such a way that only
a few qubits' interference to the transmitted qubits by an outsider
without knowing \emph{K} would cause an avalanche effect on the qubits
recovered by the receiver.}

\selectlanguage{american}%
Figure 2 shows the procedure of the QSAC. The sender first inputs
the message qubits, $\left|M\right\rangle $, and the shared key,
\emph{K,} to the QSAC algorithm. The QSAC algorithm generates check
qubits from \emph{K} and outputs a QSAC codeword which strongly entangles
the message qubits and the check qubits. Then, the sender sends the
QSAC codeword to the receiver. For verification, the receiver inputs
the received codeword and the shared key \emph{K} to the verification
algorithm to {}``extract'' the message qubits and the check qubits.
Finally, by verifying the correctness of the check qubits, the receiver
can verify whether or not the message qubits are indeed from the legitimate
participant and are without eavesdropping on modifications from an
outsider.

\begin{figure}[H]
\centering{}\includegraphics[width=0.75\textwidth]{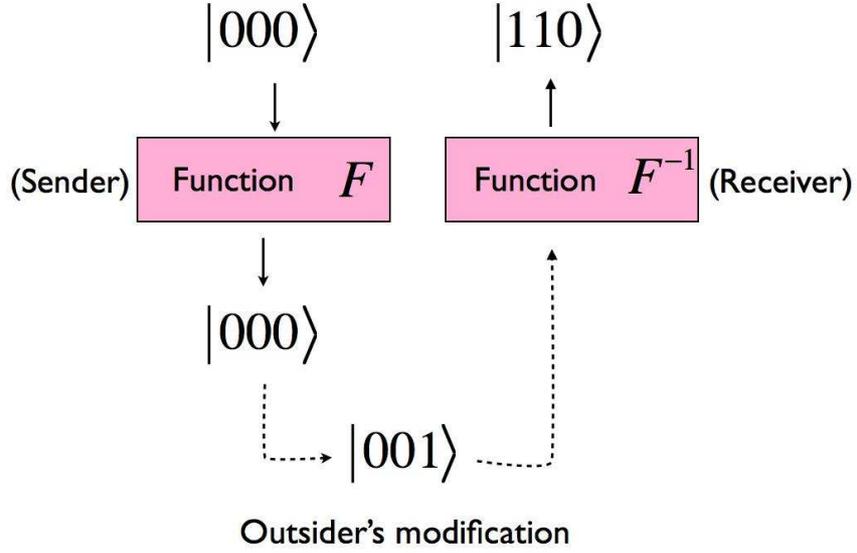}\caption{The
procedure of QSAC.}

\end{figure}

\subsection*{A. QSAC encoding:}

The following steps demonstrate the encoding of the proposed QSAC.
The inputs are the secret message qubits $\left|M\right\rangle $
of length \emph{m} and the key \emph{K }of length \emph{k} bits shared
between the sender and the receiver.
\begin{lyxlist}{00.00.0000}
\item [{Step1}] Extend the key \emph{K} to $K_{Q}$ and \foreignlanguage{english}{$K_{T}$,
via an extension function, which can be so designed as the key scheduling
in DES or AES. The lengths of }$K_{Q}$ and \foreignlanguage{english}{$K_{T}$
will be understood in the following description.}
\item [{Step2}] Transform $K_{Q}$ to a quadratic string $S_{Q}$ of length
\emph{n} and every element of $S_{Q}$ is in $\left\{ {0,1,2,3}\right\} $,
where \emph{n} is the security parameter determines the number of
check qubits. Then, a sequence of check qubits $\left|C\right\rangle =\left|S_{Q}^{1}\right\rangle \otimes\left|S_{Q}^{2}\right\rangle \otimes\cdots\otimes\left|S_{Q}^{j}\right\rangle $
can be generated according to $S_{Q}$. The elements in \{0,1,2,3\}
of $S_{Q}$ represents the states \foreignlanguage{english}{$\{\left|0\right\rangle ,\left|1\right\rangle ,\left|+\right\rangle ,\left|-\right\rangle \}$}
of check qubits, respectively. That is, if the element in $S_{Q}$
is {}``0'', then the state {}``$\left|0\right\rangle $'' is produced
and so on. For example, if $S_{Q}$ is {}``012130'', then the check
qubits should be {}``$\left|0\right\rangle \left|1\right\rangle \left|+\right\rangle \left|1\right\rangle \left|-\right\rangle \left|0\right\rangle ${}``.
\item [{Step3}] Attach the message qubits to the check qubits as $\left|\psi\right\rangle _{n+m}=\left|C\right\rangle _{n}\otimes\left|M\right\rangle _{m}$.
\item [{Step4}] Transform $K_{T}$ to a digit string $S_{T}$ of length
\emph{(n+m)} and every element of $S_{T}$ is in \foreignlanguage{english}{$\{1,2,3,\cdots,m+n\}$}\emph{.
}Perform \emph{CNOT} operations among the qubits of $\left|\psi\right\rangle _{n+m}$
according to \foreignlanguage{english}{$S_{T}$. That is,} \foreignlanguage{english}{the
index of an element in $S_{T}$} represents the index of the control
qubits in $\left|\psi\right\rangle _{n+m}$ and the corresponding
value of that element in \foreignlanguage{english}{$S_{T}$} is the
index of the target qubit in $\left|\psi\right\rangle _{n+m}$. In
other words, the CNOT operation is performed on the $i^{th}$ qubit
in $\left|\psi\right\rangle _{n+m}$, as the control qubit and the
\foreignlanguage{english}{${S_{T}^{i}}^{th}$ qubit }in $\left|\psi\right\rangle _{n+m}$\foreignlanguage{english}{,
as the target qubit.} For example, if \foreignlanguage{english}{$S_{T}=3412$,
then the sender first performs $CNOT(\left|\psi\right\rangle ^{1},\left|\psi\right\rangle ^{3})$
, then computes $CNOT(\left|\psi\right\rangle ^{2},\left|\psi\right\rangle ^{4})$
and so on.}%
\begin{table}[H]
\begin{tabular}{|c|c|c|c|c|}
\hline
Index of $S_{T}$ & 1 & 2 & 3 & 4\tabularnewline
\hline
Value of $S_{T}$ & 3 & 4 & 1 & 2\tabularnewline
\hline
Operation & \selectlanguage{english}%
$CNOT(\left|\psi\right\rangle ^{1},\left|\psi\right\rangle ^{3})$\selectlanguage{american}
 & \selectlanguage{english}%
$CNOT(\left|\psi\right\rangle ^{2},\left|\psi\right\rangle ^{4})$\selectlanguage{american}
 & \selectlanguage{english}%
$CNOT(\left|\psi\right\rangle ^{3},\left|\psi\right\rangle ^{1})$\selectlanguage{american}
 & \selectlanguage{english}%
$CNOT(\left|\psi\right\rangle ^{4},\left|\psi\right\rangle ^{2})$\selectlanguage{american}
\tabularnewline
\hline
\end{tabular}

\selectlanguage{english}%
\caption{Example of Step4 in QSAC }
\selectlanguage{american}

\end{table}

\end{lyxlist}

\subsection*{B. QSAC decoding:}

After the execution of Step4, the message qubits and the check qubits
are both strongly entangled. The verification process mainly is the
inversion of the QSAC algorithm. The inputs to the decoding function
are the received QSAC codeword as well as the shared key \emph{K.}
The following steps demonstrate the details of the verification algorithm.
\begin{lyxlist}{00.00.0000}
\item [{Step1}] The same as the Step1 in the encoding algorithm, the receiver
first produces the sub-keys $K_{Q}$ and \foreignlanguage{english}{$K_{T}$
from \emph{K}. Because of the pre-shared key \emph{K}, }the\foreignlanguage{english}{
sender and the receiver should }have the same sub-keys $K_{Q}$ and
\foreignlanguage{english}{$K_{T}$.}
\item [{Step2}] Similar to the Step4 in the encoding algorithm. The receiver
preforms the inversion of the encoding function to extract the message
qubits as well as the check qubits from the QSAC codeword and \emph{K}.
\foreignlanguage{english}{The receiver }transforms $K_{T}$ to a digit
string $S_{T}$ of length \emph{(n+m)} and every element of $S_{T}$
is in \foreignlanguage{english}{$\{1,2,3,\cdots,m+n\}$}\emph{. }\foreignlanguage{english}{The
index of $S_{T}$} represents the index of the control qubits in $\left|\psi\right\rangle _{n+m}$
and the corresponding value of the element in \foreignlanguage{english}{$S_{T}$}
is the index of the target qubit in $\left|\psi\right\rangle _{n+m}$.\foreignlanguage{english}{
It should be noted here that in the QSAC decoding, the order to perform
the CNOT operations is the reverse of that in the QSAC encoding. For
the same example, if $S_{T}=3412$, then the sender first performs
$CNOT(\left|\psi\right\rangle ^{4},\left|\psi\right\rangle ^{2})$
, then $CNOT(\left|\psi\right\rangle ^{3},\left|\psi\right\rangle ^{1})$
and so on.}%
\begin{table}[H]
\begin{tabular}{|c|c|c|c|c|}
\hline
Index of $S_{T}$ & 4 & 3 & 2 & 1\tabularnewline
\hline
Value of $S_{T}$ & 2 & 1 & 4 & 3\tabularnewline
\hline
Operation & \selectlanguage{english}%
$CNOT(\left|\psi\right\rangle ^{4},\left|\psi\right\rangle ^{2})$\selectlanguage{american}
 & \selectlanguage{english}%
$CNOT(\left|\psi\right\rangle ^{3},\left|\psi\right\rangle ^{1})$\selectlanguage{american}
 & \selectlanguage{english}%
$CNOT(\left|\psi\right\rangle ^{2},\left|\psi\right\rangle ^{4})$\selectlanguage{american}
 & \selectlanguage{english}%
$CNOT(\left|\psi\right\rangle ^{1},\left|\psi\right\rangle ^{3})$\selectlanguage{american}
\tabularnewline
\hline
\end{tabular}

\selectlanguage{english}%
\caption{Example of Step2 in verification }
\selectlanguage{american}

\end{table}

\item [{Step3}] Transform $K_{Q}$ to a quadratic string $S_{Q}$ of length
\emph{n} and every element of $S_{Q}$ is in $\left\{ {0,1,2,3}\right\} $.
With $S_{Q}$ , the receiver knows the original states of the check
qubits the sender prepared. Consequently, the receiver measures the
check qubits extracted from the QSAC codeword with the corresponding
bases. That is, if the element of $S_{q}$ is equal to 0 or 1\foreignlanguage{english}{,
then Z-basis is used}; otherwise X-basis is used.
\item [{Step4}] If the\foreignlanguage{english}{ the measurement result
of check qubits} in $S_{Q}$ is the same as the measurement result
of the check qubits recovered from the QSAC codeword, then the secret
message qubits are authenticated.\end{lyxlist}
\begin{description}
\item [{}]~
\end{description}

\subsection{Security analysis and discussions}

This section analyzes the features of the proposed QSAC including:
(1) the integrity of the secret message qubits, and (2) the originality
of the secret message qubits.

The secret message integrity ensures that the received message qubits
are not eavesdropped and modified during the time of transmission
in an open quantum channel. Conversely, if the message qubits are
eavesdropped or modified , then with a high probability the receiver
can detect the interference. To analyze the integrity, let us denote
the message qubits to be transmitted as $\left|M\right\rangle $,
the check qubits created from the key, \emph{K,} as $\left|C\right\rangle $,
and the QSAC codeword state after executing the QSAC encoding algorithm
as $\left|\psi\right\rangle $. Assume the QSAC codeword state, $\left|\psi\right\rangle $,
was modified by a malicious user into the other state $\left|\psi'\right\rangle $.
Upon receiving $\left|\psi'\right\rangle $, the receiver executes
the verification function and recovers the state to $\left|\psi''\right\rangle $.
If $\left|\psi''\right\rangle \neq\left|\psi\right\rangle $, then
the only situation that the modification passes the verification process
is: (a) a collision occurs, which means that the check qubits $\left|C''\right\rangle $
extracted from $\left|\psi''\right\rangle $ is the same as the check
qubits $\left|C\right\rangle $ produced from \emph{K}, or (b) $\left|C''\right\rangle \neq\left|C\right\rangle $,
but their measurement results are the same.

Assume that the probability for the situation (a) to occur is $p_{(a)}$.
Hence, the probability for the situation (b) to occur is $(1-p_{(a)})\times\varepsilon$,
where the\foreignlanguage{english}{ $\varepsilon$ is} the probability
for $\left|C''\right\rangle $ to have the same measurement result
as $\left|C\right\rangle $, when $\left|C''\right\rangle \neq\left|C\right\rangle $.
Therefore, the total probability for the modification to pass the
verification process when $\left|\psi''\right\rangle \neq\left|\psi\right\rangle $
is:

\[
p_{pass}=p_{(a)}+(1-p_{(a)})\times\varepsilon\]
where

\[
p_{(a)}=\frac{\left|message\; space\right|}{\left|codeword\; space\right|}\]
It is noted if the length of check qubits is large enough, then $p_{(a)}$
could approach to zero. The $p_{pass}$ thus is equal to $\varepsilon$.
Since $\varepsilon$ is the probability for \foreignlanguage{english}{$\left|C''\right\rangle $
to have the same measurement result as $\left|C\right\rangle $, where}
$\left|C''\right\rangle \neq\left|C\right\rangle $, $\varepsilon$
could be express as: $\varepsilon=\left|\left\langle C\right|\left.C''\right\rangle \right|^{2}$.
Let us assume that only one single qubits ( e.g., the $i$th check
qubit) was modified. As an example, if $\left|C\right\rangle ^{i}=\left|0\right\rangle $,
then $\left|C''\right\rangle ^{i}$ could be $\left|1\right\rangle ,\left|+\right\rangle $
or $\left|-\right\rangle $. Similarly, if $\left|C\right\rangle ^{i}=\left|1\right\rangle $,
then $\left|C''\right\rangle ^{i}=\left|0\right\rangle ,\left|+\right\rangle ,\left|-\right\rangle $
and so on. In any case, $\varepsilon=\frac{1}{3}\left(\left|\left\langle 0\right|\left.1\right\rangle \right|^{2}+\left|\left\langle 0\right|\left.+\right\rangle \right|^{2}+\left|\left\langle 0\right|\left.-\right\rangle \right|^{2}\right)=\frac{1}{3}$.
Hence if\emph{ j} check qubits are modified, then $\varepsilon=\left(\frac{1}{3}\right)^{j}$.
If the length, \emph{n}, of check qubits is large enough, then due
to the avalanche effect, $\varepsilon\doteq\left(\frac{1}{3}\right)^{\frac{n}{2}}\approx0$\foreignlanguage{english}{.}

The message originality means that the receiver can verify whether
the message is sent from the same participant whom he/she claimed
to be. To impersonate the sender, an eavesdropper must input a guessed
key \emph{K'} to the QSAC encoding algorithm to generate a codeword
$\left|\hat{\psi}\right\rangle $. If $K'\neq K$, then  the situation
for $\left|\hat{\psi}\right\rangle $ to pass the verification process
is exactly the same as the one described above, whose possibility,
$P_{Pass}$, can be ignored.

Note that the main difference between the random sampling discussion
and the QSAC is in the basic assumption for eavesdropper detection.
The former assumes that the number of check qubits {}``altered''
by an eavesdropper should be large enough, whereas the latter (QSAC)
assumes that the number of check qubits {}``set'' by the designer
is large enough. Hence, for an eavesdropper to be detected with a
high probability in the random sampling discussion, both the number
of the check qubits as well as the number of check qubits altered
by the eavesdropper should be large enough. Conversely, due to the
design of an avalanche effect on the QSAC, for a higher eavesdropper
detection rate, it requires only a large enough number of check qubits
in the QSAC.

Furthermore, to make an eavesdropper inadvertently alter enough check
qubits in the random sampling approach, the positions of those check
qubits should be unknown to the eavesdropper before these check qubits
are received by the receiver. Then, the sender and the receiver have
to communicate back and forth in an authenticated channel later to
discuss of the states, positions, and measurement results of these
check qubits to judge the existence of an eavesdropper. On the other
hand, in the QSAC, these overheads can be removed altogether. The
communication between the sender and the receiver can be simplified
to just a one-way communication.

\section{Applications}

To illustrate the usefulness of the proposed QSAC, a quantum secure
direct communication with authentication based on the Deng's two-step
QSDC\cite{key-2} is demonstrated. Instead of assuming the existence
of authenticated classical channels, we assume that the sender and
the receiver pre-share a secret key \emph{K }via some secure ways.
\begin{lyxlist}{00.00.0000}
\item [{Step~1}] The sender prepares a sequence of EPR states in $\left|\phi^{+}\right\rangle $.
The sender encodes his/her classical secret message into a message
qubits by performing one of the four operations $\left\{ I,\sigma_{x},\sigma_{z},i\sigma_{y}\right\} $
on the EPR state, which represent the two-bit classical information
$\left\{ 00,01,10,11\right\} ,$ respectively. Subsequently, the sender
encodes the message qubits to a QSAC codeword, and then sends it to
the receiver.
\item [{Step~2}] Upon receiving the QSAC codeword, the receiver decodes
the QSAC codeword and verifies the correctness of the recovered check
qubits. If the codeword is authenticated, then the receiver performs
Bell measurement on the EPR state to recover the secret message; otherwise
they abort the communication.
\end{lyxlist}
In the original two-step protocol in \cite{key-2}, the EPR pairs
should be transmitted separately in two steps in order to protect
the security of the message. Furthermore the sender and the receiver
have to perform public discussions via an authenticated classical
channel to detect the existence of eavesdroppers. However, with the
design of the QSAC, these can be done within one step and without
any tedious public discussion to detect the existence of an eavesdropper.

\section{Conclusions}

Using the inherent characteristic in quantum of being very susceptible
to be eavesdropped, this paper proposes a QSAC to protect quantum
state sequence from being eavesdropped or modified arbitrarily. As
compared to the conventionally used strategy -- the random sample
discussion, the QSAC provides efficiency in quantum sequence communication
as well as ease in the design of quantum cryptographic protocols.
The assumption of existence of authenticated classical channels in
almost all existing quantum cryptographic protocols can be removed.
Instead, a secret key is assumed to be shared between a sender and
a receiver, which is a more common and practical assumption in modern
cryptography. A QSAC-based QSDC is designed to demonstrate the usefulness
of the proposed QSAC in the design of various quantum cryptographic
protocols. On one hand, though the strongly self-entangled QSAC codeword
makes itself very difficult to be forged or interfered without detection.
On the other, it is quite susceptible to noises in a quantum channel.
Therefore, how to design a QSAC, which is robust under a noisy and
lossy quantum channel would be a very interesting future research.

\section*{Acknowledgment.}

This research is supported by the National Science Council, Taiwan,
R.O.C., under the Contract No. NSC 98-2221-E-006-097-MY3.

\end{document}